# Effects of uniaxial pressure on the quantum tunneling of magnetization in a high-symmetry Mn$_{12}$ single-molecule magnet

James H. Atkinson[1], Adeline D. Fournet[2], Lakshmi Bhaskaran[3], Yuri Myasoedov[4], Eli Zeldov[4], Enrique del Barco[1], Stephen Hill[3], George Christou[2], and Jonathan R. Friedman[5]

[1]*Department of Physics, University of Central Florida, Orlando, Florida 32816, USA*

[2]*Department of Chemistry, University of Florida, Gainesville, Florida 32611, USA*

[3]*National High Magnetic Field Laboratory and Department of Physics, Florida State University, Tallahassee, Florida 32310, USA*

[4]*Department of Condensed Matter Physics, Weizmann Institute of Science, Rehovot 7610001, Israel*

[5]*Department of Physics and Astronomy, Amherst College, Amherst, MA 01002, USA*

The symmetry of single-molecule magnets dictates their spin quantum dynamics, influencing how such systems relax via quantum tunneling of magnetization (QTM). By reducing a system's symmetry, through the application of a magnetic field or uniaxial pressure, these dynamics can be modified. We report measurements of the magnetization dynamics of a crystalline sample of the high-symmetry [Mn$_{12}$O$_{12}$(O$_2$CMe)$_{16}$(MeOH)$_4$] · MeOH single-molecule magnet as a function of uniaxial pressure applied either parallel or perpendicular to the sample's "easy" magnetization axis. At temperatures between 1.8 and 3.3 K, magnetic hysteresis loops exhibit the characteristic steplike features that signal the occurrence of QTM. After applying uniaxial pressure to the sample *in situ*, both the magnitude and field position of the QTM steps changed. The step magnitudes were observed to grow as a function of pressure in both arrangements of pressure, while pressure applied along (perpendicular to) the sample's easy axis caused the resonant-tunneling fields to increase (decrease). These observations were compared with simulations in which the system's Hamiltonian parameters were changed. From these comparisons, we determined that parallel pressure induces changes to the second-order axial anisotropy parameter as well as either the fourth-order axial or fourth-order transverse parameter, or to both. In addition, we find that pressure applied perpendicular to the easy axis induces a rhombic anisotropy $E \approx D/2000$ per kbar that can be understood as deriving from a symmetry-breaking distortion of the molecule.

I. INTRODUCTION

Single-molecule magnets (SMMs) are quantum systems that behave at low temperatures as single "giant" spins that exhibit uniquely quantum dynamics such as tunneling and interference effects. They provide an important testbed to study the boundary between the quantum and classical worlds and, in tandem, for investigation of the mechanisms of decoherence that suppress quantum dynamics. Since the initial discovery of $Mn_{12}$-acetate, many SMMs have been synthesized[1–4], including mononuclear systems[5–10], and a "wheel" built of two ferromagnetically coupled halves[11,12]. SMMs have been examined in bulk crystals, on the individual scale[13–16] and as ensembles on a surface[17,18], demonstrating a range of stable environments and applications. They are often synthesized in crystalline form, consisting of lattices of identical and well separated magnetic centers. Their relatively large net moments originate from a strong exchange interaction between intramolecular ions, producing at times a large net spin angular momentum, or from a single magnetic ion, typically possessing a sizable spin-orbit coupled angular momentum ground state.

Many properties of an SMM, such as quantum tunneling of magnetization (QTM), can be understood from its spin Hamiltonian. A simple bistable model for such a system can be represented by the "double-well" potential generated by the Hamiltonian (see *Figure 1*)

$$H = DS_z^2 - g\mu_B S_z B_z + H' \qquad (1)$$

For $D<0,$ the first term defines two energy minima corresponding to the parallel/antiparallel orientation of the spin vector *S* along an "easy" *z*-axis. The second (Zeeman) term expresses the coupling of the spin to the component of the external magnetic field parallel to the easy axis, effectively tilting the double-well potential. The last term, $H'$, contains terms that do not commute with $S_z$. Without $H'$, $S_z$ is a conserved quantity and its eigenvalues $m$ are good quantum numbers. The presence of $H'$ breaks the symmetry of the system, and therefore produces tunneling. *Figure 2* shows a diagram of the energy levels for this system (including higher-order axial anisotropy terms, e.g. $B_4^0 S_z^4$ – see below) as a function of field. When levels come into resonance, sometimes indicated with black dots in *Figure 2*, tunneling between wells takes place, leading to a marked increase in the rate of inter-well relaxation[4,19,20]. Since level "anticrossings" for one pair of levels may occur at nearly the same field as for another pair, at temperatures of a few Kelvin, tunneling may typically involve more than one pair of levels simultaneously.

The transverse anisotropy terms in $H'$ reflect the molecule's symmetry. For example, the Fe$_8$ SMM has approximate two-fold symmetry for rotation about its easy axis. This implies a rhombic transverse anisotropy:

$$H' = E\left(S_x^2 - S_y^2\right) = \frac{E\left(S_+^2 + S_-^2\right)}{2} \tag{2}$$

The spin raising and lowering operators $S_\pm$ in this transverse anisotropy couple eigenstates of $S_z$, giving rise to tunneling between such states. Moreover, since $S_\pm$ appears squared, this $H'$ imposes a selection rule: in the absence of other perturbations, tunneling can only take place between states with $m$ values that differ by integer multiples of 2. The symmetry of the molecule also allows for a fascinating geometric-phase interference effect[21–23]: tunneling between $S_z$ directions can take place along two equivalent least-action paths, one passing through the $y$ axis and the other through the $-y$ axis, allowing for interference between these paths. By applying a magnetic field along the "hard" $x$ axis, reflection symmetry in the $x$-$z$ plane for the system is preserved, maintaining the equivalence of the two least-action paths; nevertheless, the phase of the paths is modulated by the field, allowing tunneling to be dramatically suppressed when the interference between paths becomes destructive.

The Mn$_{12}$ SMM has nominal four-fold rotational symmetry in which $H'$ would have the form

$$H' = \frac{C\left(S_+^4 + S_-^4\right)}{2} \tag{3}$$

Here the fourth power of $S_\pm$ implies the tunneling selection rule $\Delta m = \pm 4n$ (integer $n$). However, in the archetypal $Mn_{12}$-acetate, the four-fold symmetry is broken by hydrogen-bonding between the $Mn_{12}$ molecules and the lattice solvent molecules in the crystal, leading to the introduction of an additional second-order transverse anisotropy in the spin Hamiltonian for most molecules in the crystal[24–26]. A geometric-phase interference effect in this system involves the interplay of a transverse field with the two different anisotropy terms[26]. In more recent years, variants of $Mn_{12}$ have been produced in which solvent effects are effectively eliminated, removing the second-order transverse anisotropy. In one of these variants, $Mn_{12}$-tBuAc, a geometric-phase interference effect can be observed[23] due to the effect of the fourth-order transverse anisotropy in which, at least at zero field, the tunneling involves four interfering equivalent paths[27,28].

In 2008, Foss-Feig and Friedman[29] predicted a novel geometric-phase interference effect in which the interference in a four-fold symmetric molecule is modulated not by an applied transverse field, but by an induced second-order symmetry-breaking anisotropy that could arise from uniaxial pressure applied along one of the hard-axis directions. This predicted interference between the four tunneling paths provides the motivation for the experiments described herein. Even in the absence of an unambiguous observation of the interference effect, we expect that a physical distortion of the molecule will induce measurable effects in the observed tunneling dynamics. Applications of pressure parallel or perpendicular to the molecular easy axis should manifest as changes to different elements of the Hamiltonian and therefore produce different behaviors as seen through QTM. In particular, pressure applied transverse to the easy axis may induce a rhombic anisotropy $E$ (cf. Eq. (2)) that affects the tunneling between certain pairs of levels. Even if the pressure is not applied precisely along one of the hard axes, a significant change in QTM should still be observed. In this case, where the pressure axis is in the $x$-$y$ plane at an angle $\varphi$ from the hard x axis, the rhombic anisotropy would be given by $\dfrac{E\left(e^{2i\varphi}S_+^2 + e^{-2i\varphi}S_-^2\right)}{2}$. In contrast, pressure applied along the easy axis should preserve the molecule's symmetry, leaving $E = 0$. Our general aim is to look for evidence of appreciable changes in the QTM behavior that can be ascribed to modifications of the molecule's anisotropy.

## II. EXPERIMENTAL SETUP

The sample under study here has the chemical formula $[Mn_{12}O_{12}(O_2CMe)_{16}(MeOH)_4]\cdot MeOH$, henceforth called "$Mn_{12}$-Me" (cf. Figure 3(b)), and is a high-symmetry analog[30,31] of the first SMM to be discovered, $[Mn_{12}O_{12}(O_2CMe)_{16}(H_2O)_4]$, a.k.a. "$Mn_{12}$-Ac". Both systems have an $S = 10$ ground state, a large energy barrier to spin reversal, $U_{eff} \sim 70$ K, and exhibit magnetic hysteresis below ~3.5 K. In crystalline samples like those examined here, the large step-like features associated with QTM are observed in the low-temperature hysteresis loops, visible every ~0.45 T or so. Since the lattice MeOH molecules form no hydrogen-bonding contacts with $Mn_{12}$ molecules, $Mn_{12}$-Me presents resonant features with minimal broadening as compared to $Mn_{12}$-Ac.

Numerous studies have been conducted in the effort to characterize pressure-induced effects in the magnetic behavior of SMMs[32–36], with pressure applied hydrostatically through the compression of a fluid medium. In contrast, our experiment was designed to deliver pressure along a particular crystalline axis of the sample. The samples studied here were small, narrow cuboid crystals with a length no greater than a millimeter. In order to aid placement and reduce the risk of crystal fracture, a method following that outlined by Campos et al.[37] was employed in which the samples were first set in epoxy (Stycast 1266) and then machined after curing in a mold. First, after removal from the mother liquor and a brief drying, the samples were placed into wet, degassed epoxy within a Teflon mold and then oriented by the application of a 4 T magnetic field held fixed for >8 hours until the completion of the curing process. The product was then machined into a small cuboid "pellet", with short dimension not much longer than the length of the sample itself, such that one of the flat faces of the crystal was close to one of the flat faces of the pellet.

The cured epoxy pellet containing the embedded crystal was placed within a "bracket" designed to hold a Hall Bar magnetometer and G10 "fingers" that served to deliver pressure to two opposing faces of the crystal. The sample was aligned with the active area of the Hall sensor to achieve good coupling. Figure 3 (c and d) schematically illustrates the low-temperature portion of the apparatus. The bracket containing the embedded sample and sensor was inserted into a stainless-steel cell at the bottom of an apparatus designed to deliver pressure to the sample from a pneumatic piston outside of the cryostat. Pressure was applied by supplying compressed nitrogen gas through a regulator to the piston. The pressure applied to the sample was calculated using the dimensions of the piston and the cross section of the epoxy pellet containing the sample. The apparatus was essentially identical to that used in a previous experiment [38].

Two Quantum Design PPMS systems were used for measurements[39], each with similar instrumentation but different orientations of the magnetic field axis. For one, the field axis was aligned with the bore of the sample space. For the other, the field axis was perpendicular to the sample-space bore. In the first system, the sample was aligned such that its easy axis (collinear with the long axis of the crystal) was parallel to the magnetic field, which was also parallel to the pressure axis $P$. This is the "parallel pressure" configuration (*Figure 3*(c)-left). In the second system, the easy axis and magnetic field are again collinear, but $P$ is oriented orthogonally to both. This is the "perpendicular pressure" configuration *(Figure 3*(c)-right). In this latter configuration, the pressure was applied along the "hard" anisotropy axis of the sample [40].

Hall resistance from our sensors was measured using a lock-in amplifier. The applied field was swept at rates of 1.1 mT/s and 19.9 mT/s in the parallel-pressure configuration and 8 mT/s in the perpendicular configuration. Measurements were performed across a range of temperatures from 1.8 to 3.3 K, as read from a calibrated thermometer placed on the back of the Hall bar in the parallel case, and according to the PPMS system thermometer at the sample-space position in the perpendicular case. After each adjustment of the pressure, hysteresis data was collected at several temperatures before moving on to the next pressure. For the parallel experiment, the sequence of pressures began with ambient pressure ("zero" pressure in the piston) and was then followed by pressures of 0.55, 1.1, 1.64, 0.27, 0.82, 1.37, and 1.92 kbar. For the perpendicular experiment, the sequence was 0.41, 0.82, 1.23, 1.64, 2.05, and finally 0 kbar. The upper limit of the pressures was constrained by the expected compressive strength of the epoxy at low temperatures. The same pellet/crystal was measured in both experiments; it was rotated between experiments to accommodate the different pressure configurations.

## III. RESULTS

*Figure 3*(a) shows hysteresis loops for several temperatures acquired in the perpendicular pressure configuration while at ambient pressure. The measured hysteresis curves show clear QTM steps that can be labelled by the conventional resonance-numbering system, as indicated in *Figure 3*(a). Relaxation associated with the resonance condition for a single resonant pair of levels is not discernible due to the broadening present at these temperatures.

From the hysteresis data, one can determine the spin relaxation rate $\Gamma$:

$$\Gamma \equiv \frac{-1}{M - M_{eq}} \frac{dM}{dt} = \frac{-\dot{B}}{M - M_{eq}} \frac{dM}{dB}, \tag{4}$$

where $M$ represents the instantaneous magnetization of the sample and $M_{eq}$ the equilibrium magnetization. $M_{eq}$ can be calculated using Boltzmann statistics, and converges to the magnetization of a "saturated" sample $M_{sat}$ at sufficiently large fields. *Figure 4* shows the $B$ dependence of $\Gamma$ for the k = 2 and 4 resonances for several values of pressure applied in the parallel and perpendicular configurations, respectively. The data show that pressure in both configurations enhances the relaxation rate ($\alpha$, which characterizes the rate for a given resonance as described below) at tunneling resonances. In addition, the resonance fields ($\beta$, which characterizes the resonance's center field) appear to shift, with opposite trends between the two different orientations of pressure: the resonance field shifts to larger values with increasing pressure in the parallel configuration and towards lower fields in the perpendicular case (see the insets of *Figure 4,* as well as *Figure 5* below). The results imply no significant permanent deformation of the sample, given that the data follow the same general pressure dependence, regardless of the fact that pressure was not changed monotonically.

To further analyze the data, we assume that $\Gamma$ can reasonably be expected to follow the convoluted lineshapes of the levels involved in tunneling. To wit,

$$\Gamma = f(B) \tag{5}$$

where *f(B)* is the lineshape function associated with the tunnel resonance, such as a Lorentzian,

$$f_L(B) = \frac{\alpha \frac{w}{2}}{\pi\left[\left(\frac{w}{2}\right)^2 + (B-\beta)^2\right]}, \text{ or Gaussian, } f_G(B) = \frac{\alpha}{w\sqrt{2\pi}} e^{-\frac{(B-\beta)^2}{2w^2}}, \text{ where } w \text{ is the linewidth, } \beta \text{ is}$$

the resonance center and $\alpha$ characterizes the "strength" of the tunneling resonance (the amplitude or area of the peak). In fitting our data, we employed a "pseudo-Voigt profile" function [41]:

$$f(B) = \xi f_L(B) + (1-\xi) f_G(B), \tag{6}$$

with $0 < \xi < 1$ a free parameter, giving the relative weights for the Lorentzian and Gaussian functions. Fitting using this expression yields values for $\alpha$, $\beta$ and $w$ for each QTM step, with a single value for each parameter shared between the Lorentzian and Gaussian components. Prior to fitting, a correction was made to account for the contribution of the mean internal dipole field to the effective local field as seen by the spins [42], with a value of 22.5 mT used for a fully magnetized sample.

Due to the significant broadening we observe, not only are individual level pairs within a given step indistinct, but it is moreover impossible to completely separate the steps and attribute all the observed relaxation to a particular *k*, as appreciable relaxation is often found at field values between the step "centers" (e.g. between $k = 1$ and 2), albeit at a slower rate. The fitting was performed with this in mind by allowing the possibility of overlap between neighboring fit functions. A range of data (spanning ~4500 Oe) centered about the "target" step was selected to which three functions were fit, two of which represented contributions from "neighbor" resonances on either side of the target step. In the first step of an iterative process, the function fit to the "target" step contained free parameters for $α$, $w$, and $β$, whereas the associated parameters for the "neighbor" steps were held fixed. Subsequently, one of the other two "neighbor" functions was "unlocked" and the process repeated with the other two held fixed. This process was repeated many times until the parameter estimates reached stable values.

Some of the extracted values for parameters $α$ and $β$ are shown in the insets of *Figure 4*, where they are plotted as a function of pressure. Linear fits were made to the pressure dependence for each set of data identified by its associated step $k$, temperature, sweep rate, and orientation of pressure. The slopes of the fits were normalized by the zero-pressure intercept to generate $(∂α/∂P)/α(0)$ and $(∂β/∂P)/β(0)$. Extracted values for the resonance widths, $w$, show no clear trend and/or large error bars and, as such, we omit their analysis. (Analysis of the estimates for $\xi$ also did not reveal any clear trends with pressure.)

*Figure 5* shows the collected data for these two quantities (open squares), delineated by the resonance label $k$. Generally, we find that when the pressure was parallel to the easy axis, the resonances grew noticeably larger ($(∂α/∂P)/α(0) \lesssim 10\%$/kbar) and shifted towards higher field ($(∂β/∂P)/β(0) \lesssim 0.3\%$/kbar) as the pressure was increased. For pressure perpendicular to the easy axis, the magnitudes again increased ($(∂α/∂P)/α(0) \lesssim 5\%$/kbar), but the steps instead shifted towards lower field with increasing pressure ($(∂β/∂P)/β(0) \lesssim -0.4\%$/kbar). Previous work performed with this apparatus in which a sample of the lower-symmetry $Mn_{12}$-Ac was placed in a parallel-pressure configuration exhibited an increase in the resonance positions of 0.11-0.14%/kbar[38].

## IV. ANALYSIS & INTERPRETATION

To understand the observed pressure dependence, we calculated how changes to different anisotropy parameters in the molecule's spin Hamiltonian would affect the magnetization relaxation rate. In general, changes in the axial parameters will act to shift eigenstates up or down in energy, affecting their Boltzmann populations for a given temperature and changing the fields at which levels in opposite wells cross (the resonance fields). In contrast, tuning the transverse anisotropy parameters can result in pronounced changes in the expected tunneling rate while leaving the resonance fields largely unchanged.

In the absence of a symmetry-breaking perturbation such as perpendicularly applied pressure or field, the spin Hamiltonian for $Mn_{12}$-Me can be taken to be

$$H = DS_z^2 + AS_z^4 + FS_z^6 + B_4^4 O_4^4 + B_6^4 O_6^4 - g\mu_B \vec{S} \cdot \vec{B}, \qquad (7)$$

where $O_4^4 = (S_+^4 + S_-^4)/2$, and $O_6^4 = \{11S_z^2 - S(S+1) - 38, (S_+^4 + S_-^4)\}$ are Stevens operators, with the curly brackets representing the anticommutator operation: $\{a,b\} = ab + ba$. We use the ambient-pressure values for the anisotropy parameters determined previously for a similar high-symmetry $Mn_{12}$ molecule[43]: $D = -0.567$ K, $A = -6.91 \times 10^{-4}$ K, $F = -3.31 \times 10^{-6}$ K, $B_4^4 = 2.88 \times 10^{-5}$ K, and $B_6^4 = -1.44 \times 10^{-7}$ K.

For initial characterization of the parallel-pressure results, we performed regression fitting in conjunction with a Hamiltonian without transverse $B_4^4$ and $B_6^4$ terms:

$$H = DS_z^2 + AS_z^4 + FS_z^6 - g\mu_B \vec{S} \cdot \vec{B}. \qquad (8)$$

In the absence of a transverse field, this Hamiltonian contains no terms that would permit tunneling (i.e. there are no off-diagonal terms to mix different spin eigenstates). *Figure 2* shows the energy levels calculated using this restricted Hamiltonian superimposed on relaxation-rate data taken under ambient pressure for several temperatures, as indicated. We focus on resonances $k = 1, 2, 3,$ and 4, for which we have data at several temperatures in both the parallel and perpendicular cases. We assume that, for a given $k$, substantial tunneling takes place via no more than two resonances (level anticrossings), a paradigm we refer to as the "two-state approximation". We choose levels that match closely with observed resonance positions, as indicated in *Figure 2*, and describe the total transition amplitude as a Boltzmann-weighted linear sum of contributions from the two levels, i.e. $α_{Tot} = α_1exp(-E_1/k_BT) + α_2exp(-E_2/k_BT)$, where the subscripts refer to levels with values $m_1$ and $m_2$, as given in Table I.

Table I – List of initial (*m*) and final (*m'*) states used in modelling tunneling in the two-state approximation for each resonance $k = -(m + m')$.

| Resonance $k$ | $m_1, m_1'$ | $m_2, m_2'$ |
|---|---|---|
| 1 | -7, 6 | -6, 5 |
| 2 | -7, 5 | -6, 4 |
| 3 | -7, 4 | -6, 3 |
| 4 | -8, 4 | -7, 3 |

By taking estimates for the amplitudes $α_1$ and $α_2$ to be constant as a function of pressure, we can interpret changes in the observed resonance magnitudes as alterations of the Boltzmann populations induced by changes in *D* and *A*. The sixth order constant *F* was held constant in order to limit the number of free parameters and because good agreement could be found without letting it vary. Fitting the temperature dependence of the zero-pressure intercepts from the linear fits of the amplitude *α* generated initial estimates for $α_1$ and $α_2$. Simultaneously, we fit the temperature dependence of the ambient *β* data to an expression that approximates the resonance field as a sum of the two individual resonance fields of the contributing levels, weighted by their Boltzmann populations and transition amplitudes $α_1$ and $α_2$, i.e. $β_{Tot} = β_1α_1exp(-E_1/k_BT) + β_2α_2exp(-E_2/k_BT)$. In this way, we could produce estimates for changes in the Hamiltonian parameters by fitting to the parallel data as a function of pressure and temperature and letting *D* and *A* vary. The results of this fitting give values of $(∂D/∂P)/D = 0.56\%$/kbar and $(∂A/∂P)/A = -6.4\%$/kbar.

To analyze the data more fully in both the parallel and perpendicular configurations, we employed a master-equation approach to calculate the relaxation rate $Γ$ by accounting for spin-phonon interactions [23,44,45]:

$$\frac{dp_i}{dt} = \sum_{\substack{j=1 \\ i \neq j}}^{21} -(\gamma_{ij}^{(1)} + \gamma_{ij}^{(2)})p_i + \left(\gamma_{ji}^{(1)} + \gamma_{ji}^{(2)}\right)p_j, \tag{9}$$

where the $p_i$ are the populations of the system's energy eigenstates (diagonal elements of the density matrix). In this so-called secular approximation, eigenstates were calculated using a Hamiltonian that included transverse anisotropy terms up to sixth order:

$$H = DS_z^2 + AS_z^4 + FS_z^6 + E\left(S_x^2 - S_y^2\right) + B_4^4 O_4^4 + B_6^4 O_6^4 - g\mu_B \vec{S}\cdot\vec{B}, \quad (10)$$

where in the absence of applied pressure the coefficients have the "ambient" values given above, and $E$ has an ambient value of zero. We consider only diagonal elements in the density matrix to conserve computational resources; this reduces the applicability of the model near the resonance condition where off-diagonal elements might play an important role. Nonetheless, this model allows us to discriminate the effects of varying different anisotropy parameters. The spin-phonon transition rates are given by [44,45]

$$\gamma_{ij}^{(1)} = \frac{D^2}{24\pi\rho c_s^5 \hbar^4} \left|s_{ij}^{(1)}\right|^2 \Delta_{ij}^3 N(\Delta_{ij}), \quad (11)$$

$$\gamma_{ij}^{(2)} = g_2 \frac{D^2}{36\pi\rho c_s^5 \hbar^4} \left|s_{ij}^{(2)}\right|^2 \Delta_{ij}^3 N(\Delta_{ij}), \quad (12)$$

where $\rho$ is the mass density of the sample, $c_s$ is the speed of sound in the sample, $s_{ij}^{(1)} = \langle i|\{S_x, S_z\}|j\rangle$, $s_{ij}^{(2)} = \langle i|S_x^2 - S_y^2|j\rangle$, $\Delta_{ij}$ is the energy difference between levels $i$ and $j$, $N(\Delta_{ij}) = 1/[exp(-\Delta_{ij}/k_B T)-1]$ is the phonon thermal-distribution function, and $g_2 = 1.125$ is a fitting parameter. The rate $\Gamma_{sim}$ is taken as the smallest non-zero eigenvalue of the matrix generated from Eq. (9)[23]. In our simulation, we assumed a constant field $H_x = 10$ mT in order to represent the effects of internal (dipole or hyperfine) transverse fields that have been demonstrated to play a significant role in the magnitudes of QTM resonances[46].

In order to extract estimates for resonance fields from calculations of $\Gamma_{sim}$ we used a weighted integral over a field range centered around the resonances of interest (such that it included all "peaks" associated with a given resonance number $k$):

$$\beta_{sim} = \frac{\int_{B_i}^{B_f} \Gamma(x) x \, dx}{\int_{B_i}^{B_f} \Gamma(x) \, dx} \tag{13}$$

in which the bounds $B_i$ and $B_f$ about a resonance $k$ are taken to be $0.45k \pm 0.15$ T. The spin-phonon rate simulation produces data in which $\Gamma_{sim}$ is small except near a resonance condition and, as such, Eq. (13) is essentially an average of the peaks' abscissas weighted by their areas. The amplitude $\alpha_{sim}$ of a resonance was simply taken as the total area under the peaks for the same bounds. To conserve computational resources, contributions from specific resonances were omitted when we could safely assume they were small, i.e. resonances closest to the ground state that were calculated to occur at fields far from where relaxation was observed.

Inserting the estimated values for $(\partial D/\partial P)/D$ and $(\partial A/\partial P)/A$ extracted from the parallel, two-state approximation scheme outlined above into the master-equation simulations reproduces changes in the calculated amplitudes and resonance fields (i.e. as estimated by Eq. (13), represented by the red circles in Figure 5) similar to those found from the two-state approximation (i.e. via the Boltzmann weighted sum that determines $\beta_{Tot}$), confirming the consistency of the two models.

The red line in *Figure 6*(a) shows the simulated relaxation rate as a function of field for the $k = 2$ resonance generated using modified values for the anisotropy parameters, as indicated, including the values for $(\partial D/\partial P)/D$ and $(\partial A/\partial P)/A$ discussed above. The dashed blue line corresponds to an alternative model in which $D$ and the tetragonal transverse anisotropy term $B_4^4$ were simultaneously varied, while A is held fixed at its ambient value. In this scenario, a slight increase in $D$ of 0.042% ($\pm$0.015) and an increase in $B_4^4$ of 4.3% ($\pm$0.36) create a small increase in the step center field and an enhancement of the tunneling rate in the vicinity of all resonances; in this model, the effect of $\partial D/\partial P$ is almost negligible, as most of the observed effects can be attributed to a change in $B_4^4$. A model in which only $D$ is allowed to vary fails to adequately reproduce the observed behavior.

The green lines in *Figure 6*(a) show the result of introducing a rhombic term $|E|< 0.3$ mK ($\approx$ $D$/2000), oriented along one of the hard axes, without changing other parameters. Contrary to the effects produced through changes in $D$, $A$ or $B_4^4$, the introduction of the rhombic term "opens" certain resonances (e.g. the one near 0.865 T – see also Figure 7) while leaving the resonance positions largely unchanged. The most dramatic effects appear at positions corresponding to excited states near the top of the barrier, which (for the same resonance $k$) are always closer to zero field than lower-energy states (cf. *Figure 2*) since both $D$ and $A$ are negative. This implies that an average over these resonances would shift towards lower field as the rhombic component is increased, a conclusion borne out by the estimates of $\beta_{sim}$, as illustrated in *Figure 6*(b), which shows the results of simulations for three experimentally relevant temperatures. One can see that the resonance field decreases with increasing temperature as a greater portion of the tunneling takes place through higher excited states.

The $\beta_{sim}$ and $\alpha_{sim}$ dependences were calculated while either $D$ and $A$ or $D$ and $B_4^4$ were varied (for the parallel configuration) or an $E$ term was introduced (for the perpendicular case). These scenarios were simulated for three temperatures, 2.1 K, 2.7 K, and 3.3 K. These generated data qualitatively similar to that shown in *Figure* (b), with approximately linear trends evident. Lines were fit to these calculations to determine the dependence of $\beta_{sim}$ and $\alpha_{sim}$ on $D$ and $A$, $D$ and $B_4^4$, or $E$. *Figure 5* shows the estimated changes in the amplitude $(\partial\alpha/\partial P)/\alpha(0)$ and center positions $(\partial\beta/\partial P)/\beta(0)$ resulting from our simulations, plotted on top of the equivalent estimated changes extracted from our data. The simulations correspond to changes in the anisotropy parameters of $(\partial D/\partial P)/D(0) = 0.56$ %/kbar and $(\partial A/\partial P)/A(0) = -6.4$ %/kbar (red points, generated from the two-state methodology outlined above and representing the case of $\boldsymbol{P} \parallel z$), or $(\partial D/\partial P)/D(0) = 0.042$ %/kbar and $(\partial B_4^4/\partial P)/B_4^4(0) = 4.3$ %/kbar (blue points, also representing $\boldsymbol{P} \parallel z$), or $\partial E/\partial P = -0.29$ mK/kbar (for $\boldsymbol{P} \perp z$). In all cases, the simulations show good agreement with the trends observed in the experimental data, matching the directions and the relative sizes of the changes of the shifts in both parameters. The scatter present in the values derived from the experimental data (open squares in *Figure 5*) would likely accommodate a range of scenarios involving different changes to the anisotropy parameters, such as the alteration/introduction of additional higher-order Stevens operators, but we believe the estimates we have proposed represent the simplest scenarios capable of reproducing results similar to our experimental data.

## V. DISCUSSION

For the parallel-pressure case, we find that pressure induces changes consistent with two possible models: a reduction in the fourth-order axial anisotropy $A$ and a slight increase in the second-order axial anisotropy $D$, or an increase in the transverse tetragonal anisotropy $B_4^4$ and a very slight increase in $D$. In the former case, the anisotropy changes combine to produce an increase in the resonance positions while simultaneously reducing the energies of the excited levels and thereby increasing the relaxation rate. In the latter case, the increased relaxation is due to an increase in the tunneling produced by the larger value of $B_4^4$. As we discuss below, these variations in $A$ or $B_4^4$ may be due to pressure-induced modulations of the intramolecular exchange interactions coupling the constituent manganese ions[34,47–49], caused by changes in the structure of the $Mn_{12}$ molecule.

The results here stand in some contrast to those produced in an analogous experiment on $Mn_{12}$-Ac in which the effect of parallel pressure was to shift the resonances to higher field by an amount which falls at the low end of the range of values per kilobar of that seen here. The difference may be due to the solvent disorder in $Mn_{12}$-Ac possibly dominating changes in the molecular anisotropy when compared with the effect of pressure. To wit, if, in the parallel configuration, the pressure serves to increase $B_4^4$, $Mn_{12}$-Ac would be less susceptible to the effects of pressure since tunneling is already determined by the substantial value of $E$ induced by the solvent-bonding effect. Thus, the difference in the effect of parallel pressure for the two systems would favor the model that pressure is increasing $B_4^4$, over the one with decreasing $A$.

*Figure 7* illustrates the effect that introduction of a rhombic term has on the simulated relaxation rate for different resonances at the steps $k = 2$ and 3. For the $k = 2$ step, resonances that represent $\Delta m$ values that are integer multiples of the order of the dominant transverse symmetry (in this case four, as the molecule has tetragonal symmetry) are largely unaffected, as can be seen in the peak labelled by $m = -5 \rightarrow +3$, for which $\Delta m = 8$. However, transitions for which $\Delta m = \pm 2n$ ($n$ odd), such as the $m = -4 \rightarrow +2$ peak, show large increases in associated relaxation rates as the introduction of rhombic transverse anisotropy opens up a new pathway to tunneling that is otherwise largely forbidden by the $\Delta m = 4n$ selection rule. A similar mechanism is at work in the resonances at odd steps, as some transitions allowed by the combination of dominant anisotropy terms and a transverse field remain largely unaffected by the introduction of a rhombic term (see the peaks labelled $m = -4 \rightarrow +1$ and $m = -6 \rightarrow +3$ in *Figure 7*(b)), while other resonances become somewhat more intense when a significant $E$ term is present (such as the $m = -5 \rightarrow +2$ peak). Each peak is labelled in terms of the transverse Hamiltonian terms that, to leading order of perturbation theory, produce the tunneling for that peak. For example, the $m = -5 \rightarrow +2$ peak is labelled "$B_4^4 + E + B_T$", indicating that this transition requires third-order perturbation theory involving $B_4^4$ to first order ($\Delta m = 4$), $E$ to first order ($\Delta m = 2$) and a transverse field, $B_T$, to first order ($\Delta m = 1$). For $E = 0$, the transition is still possible in the same order ($B_4^4$ twice and $B_T$ once), and so $E$ will enhance this transition only when the two processes have comparable magnitude. We find that transitions occurring closer to the top of the potential barrier (i.e. where $\Delta m$ is less) typically show larger increases in the calculated relaxation rate for a given change in $E$, as they require lower order of perturbation theory to mix the states involved.

Our results suggest that at the highest pressure we applied (~2 kbar), we induced a transverse anisotropy of approximately $E = D/975 \approx 0.6$ mK. Foss-Feig and Friedman[29] showed that for $Mn_{12}$, the introduction of a rhombic component $E \approx 15$ mK is capable of inducing strong destructive Berry-phase interference, resulting in quenching of the dominant tunneling mechanism. Unfortunately, the pressure our apparatus can apply is limited by the compressive strength of the epoxy encasing the crystalline sample. Tests of the epoxy cooled in liquid nitrogen showed that it will fracture before ~5 kbar. Thus, substantial changes to the experimental apparatus would be necessary to observe the Berry-phase interference in this SMM.

The spin dynamics in polynuclear SMMs arise from a competition between the single-ion anisotropies originating from spin-orbit coupling, and the inter-ion exchange interactions. When the latter are substantially stronger, an SMM can be well described with a giant-spin Hamiltonian up to second order in the axial spin operator ($DS_z^2$), assuming a high molecular symmetry. However, for inter-ion exchange couplings comparable to the single-ion anisotropies, higher-order axial ($AS_z^4$) and/or transverse ($B_4^4 O_4^4$) corrections to the leading anisotropy term need to be included in the giant-spin approximation. The nature of high-order anisotropy corrections in the giant-spin approximation, and their relation to inter-ion exchange couplings, have been well established and discussed at length in previous work[43,47,50–55]. In brief, these corrections arise from mixing of different spin multiplets separated by an energy proportional to the exchange interaction strength. In the limit of infinite exchange, quantum spin fluctuations vanish completely (i.e. the molecular spin is well defined), and the allowed terms in the Hamiltonian are limited to second order. Note that, although the exchange is not directly responsible for the higher-order corrections to spin-orbit coupling, its magnitude dictates the degree of spin mixing, consequently determining the appearance of higher-order terms. In this way, pressure-induced changes to the inter-ion exchange couplings, most likely via mechanical distortion of the respective Mn-O-Mn superexchange bonds and angles, provide a highly plausible explanation for the observed changes in tunneling rates. However, pressure-induced changes in the orientations of the single-ion anisotropy tensors with respect to the molecular symmetry axis provide an alternative mechanism for the observed changes in tunneling rates. Indeed, in the case of purely axial single-ion anisotropy (single-ion $e = 0$), tilts of the corresponding tensors away from the molecular symmetry axis would be the only viable source of transverse anisotropy (provided moderate exchange energies). In that case, variations of these tilts (*e.g.*, by uniaxial pressure) could have a sizeable effect on the tunneling rates and positions of QTM resonances. However, the single-ion anisotropies of the Mn ions are typically rather rhombic (both *d* and *e* terms).[46,47] In this case, the molecular transverse anisotropy results from projections of both *d* and *e* onto the molecular transverse symmetry plane. Hence, small changes of the orientations of the local anisotropy tensors induced by pressure will produce less pronounced changes in the tunneling rates and resonance positions, although such effects cannot be completely ignored. Most likely, the effects observed in this work have contributions from both mechanisms, i.e., pressure-induced variations of both the inter-ion exchange and the orientations of the local Mn anisotropy tensors. The relatively weak changes in relaxation obtained within the range of pressures employed in this work do not allow us to discern the relative role of the two mechanisms.

Fitting the parallel-pressure data could be improved by allowing all three parameters (*D, A* & $B_4^4$) to vary independently. In fact, one might expect *A* and $B_4^4$ to follow similar behavior, given their relationship to intra-molecular exchange. However, fitting using a model that requires these two parameters to change proportionately yielded unsatisfactory results. Given the precision of the data presented here, a three-parameter fit would not be illuminating. Experiments performed either under different conditions, e.g., lower temperatures, or employing other methods, such as electron-spin resonance spectroscopy or X-ray diffractometry, may be able to discriminate between the different models and reveal finer details of the mechanisms involved. However, none of these approaches are currently compatible with the uniaxial pressure apparatus employed in this study. On the other hand, the interpretation that perpendicular pressure induces an *E* term, increasing tunneling by breaking the molecule's symmetry, appears relatively unambiguous and our results provide the first evidence of such a uniaxial-pressure-induced effect in an SMM.

In summary, we have demonstrated that uniaxial pressure applied parallel or perpendicular to the easy axis of a crystalline sample of $Mn_{12}$-Me alters the process of QTM relaxation with behavior that depends on the direction of the pressure. The observations can be explained as arising from pressure-induced changes to the molecule's anisotropy parameters.

ACKNOWLEDGEMENTS

We thank Professor Mark Meisel and his group members for supplying samples of the epoxy for preliminary testing. D. Krause, R. Cann, and J. Kubasek all provided valuable insight and precise work in the manufacture of the apparatus and machining of the epoxy pellets. J. Marbey, D. Komijani, and S. Greer kindly provided technical assistance, transportation, and generous hospitality. Support for this work was provided by the U.S. National Science Foundation under Grants No. DMR-1006519 and No. DMR-1310135. J.H.A. and E.d.B. acknowledge support from the U.S. National Science Foundation under GrantsNo. DMR-1503627 and No.DMR-1630174. A.D.F. and G.C. acknowledge support from the U.S. National Science Foundation under Grant No. CHE-1565664. S.H. and L.B acknowledge support from the U.S. National Science Foundation under Grants No. DMR-1309463 and No. DMR- 1610226. J.R.F. acknowledges the support of the Amherst College Senior Sabbatical Fellowship Program, funded in part by the H. Axel Schupf '57 Fund for Intellectual Life. The National High Magnetic Field Laboratory is supported by the State of Florida, and work performed there was supported through the U.S. National Science Foundation Cooperative Agreement No. DMR-1157490.

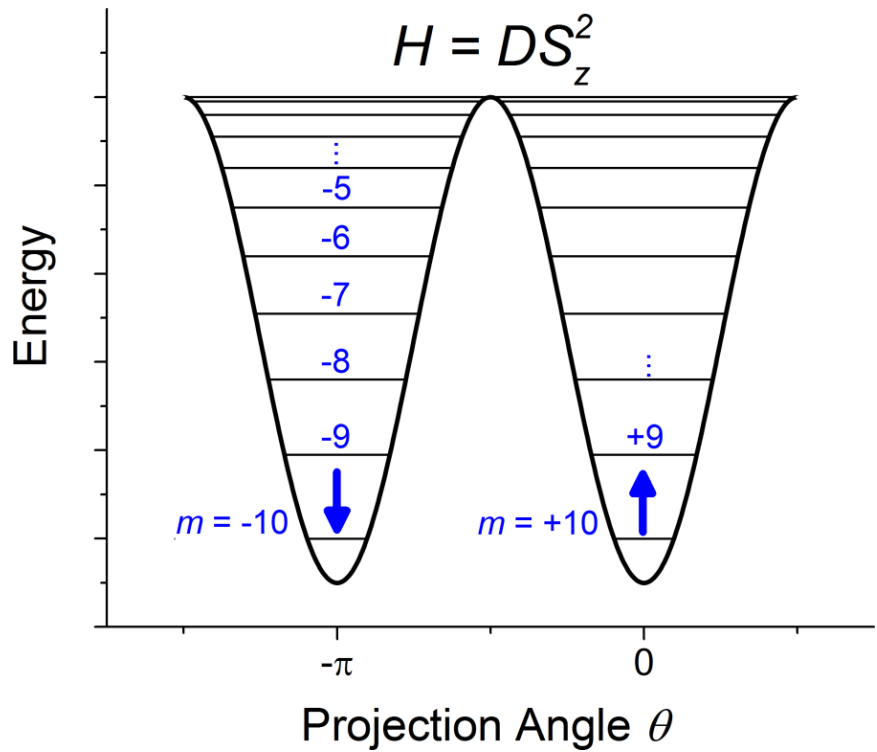

*Figure* 1 Double-well potential generated from the second-order anisotropy term in Eq. (1) for an $S = 10$ system. The horizontal axis is the classical projection angle of the spin vector along the "easy" ($z$) *axis.* The eigenstates are labelled by their associated magnetic quantum number $m$.

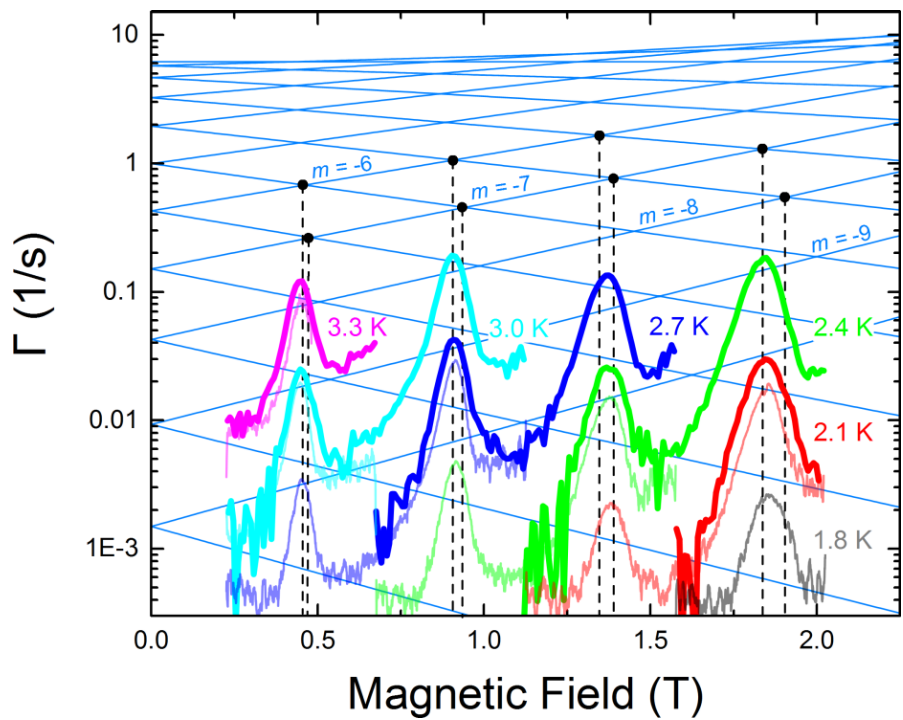

*Figure* 2 Logarithmic plot of relaxation rate $\Gamma$ from ambient pressure data taken in the parallel configuration, overlaid on an energy level diagram generated from the $Mn_{12}$-Me Hamiltonian. The thick lines correspond to data acquired at a sweep rate of 19.9 mT/s, while the thin lines are smoothed curves generated from data acquired using a sweep rate of 1.1 mT/s. The contributing resonances assumed in the two-state approximation (see text and Table ) are indicated by the black dots, with the dashed lines aligned with the corresponding resonance fields.

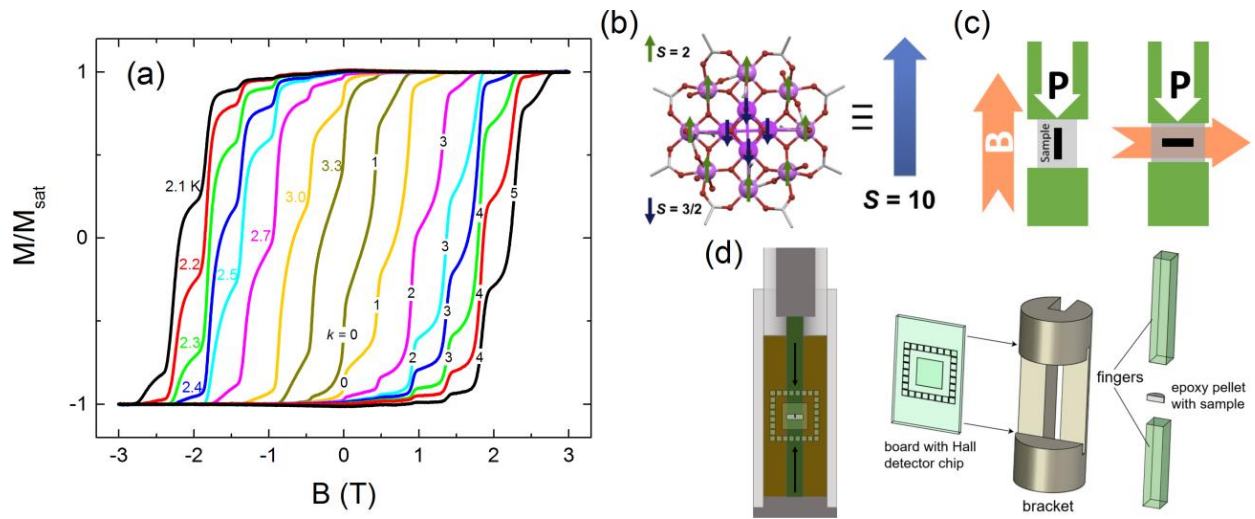

*Figure 3* (a) Temperature dependence of hysteresis loops acquired under ambient conditions in the perpendicular-pressure configuration at a longitudinal-field sweep rate of 8 mT/s, labelled by their resonance number $k = -(m + m')$. (b) Schematic of the molecular core of $Mn_{12}$-Me, from ref. [30], with the dark blue/green arrows illustrating relative spin orientations of the individual Mn ions and a large blue arrow representing a collective "giant spin". (c) Diagram showing the relative orientations of the sample and magnetic field within the pressure apparatus. The sample's easy magnetization axis is along the long dimension of the sample (black rectangle). (d) Detailed schematic of the high-pressure cell within the low-temperature portion of the apparatus showing the arrangement of the Hall-bar sensor and the elements that deliver pressure to the sample, adapted from ref. [38]. Note that the shape of the epoxy "pellet" used in this experiment is close to a cuboid.

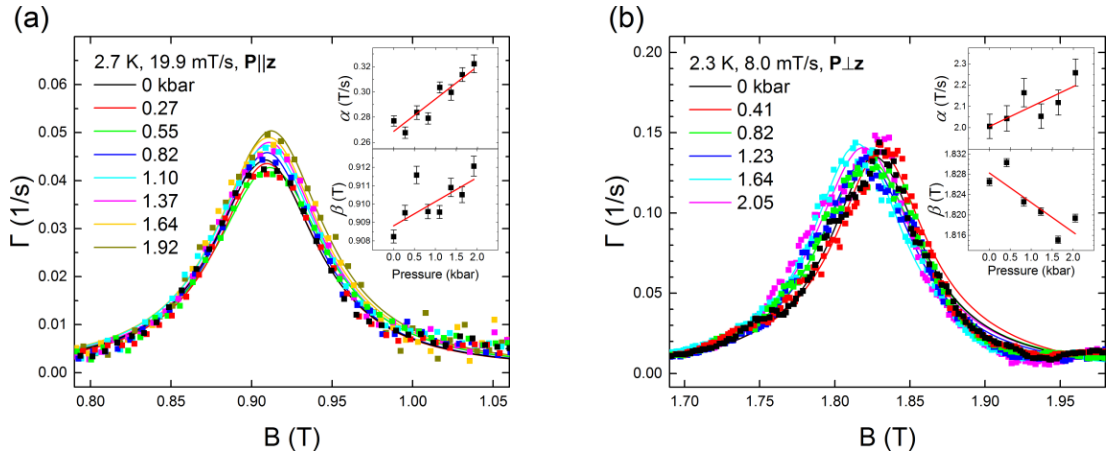

*Figure 4* Extracted relaxation rate for the (a) $k = 2$ and (b) $k = 4$ resonances. The $k = 2$ data were acquired at 2.7 K with a sweep rate of 19.9 mT/s in the parallel configuration. The $k = 4$ data shown represents an adjacent-average smoothing of actual data acquired in the perpendicular configuration, which were taken at 2.3 K with a sweep rate of 8.0 mT/s. The solid lines are curve fits to the data. The insets show the pressure dependence of the fit parameters $\alpha$ and $\beta$, with linear fits (red lines).

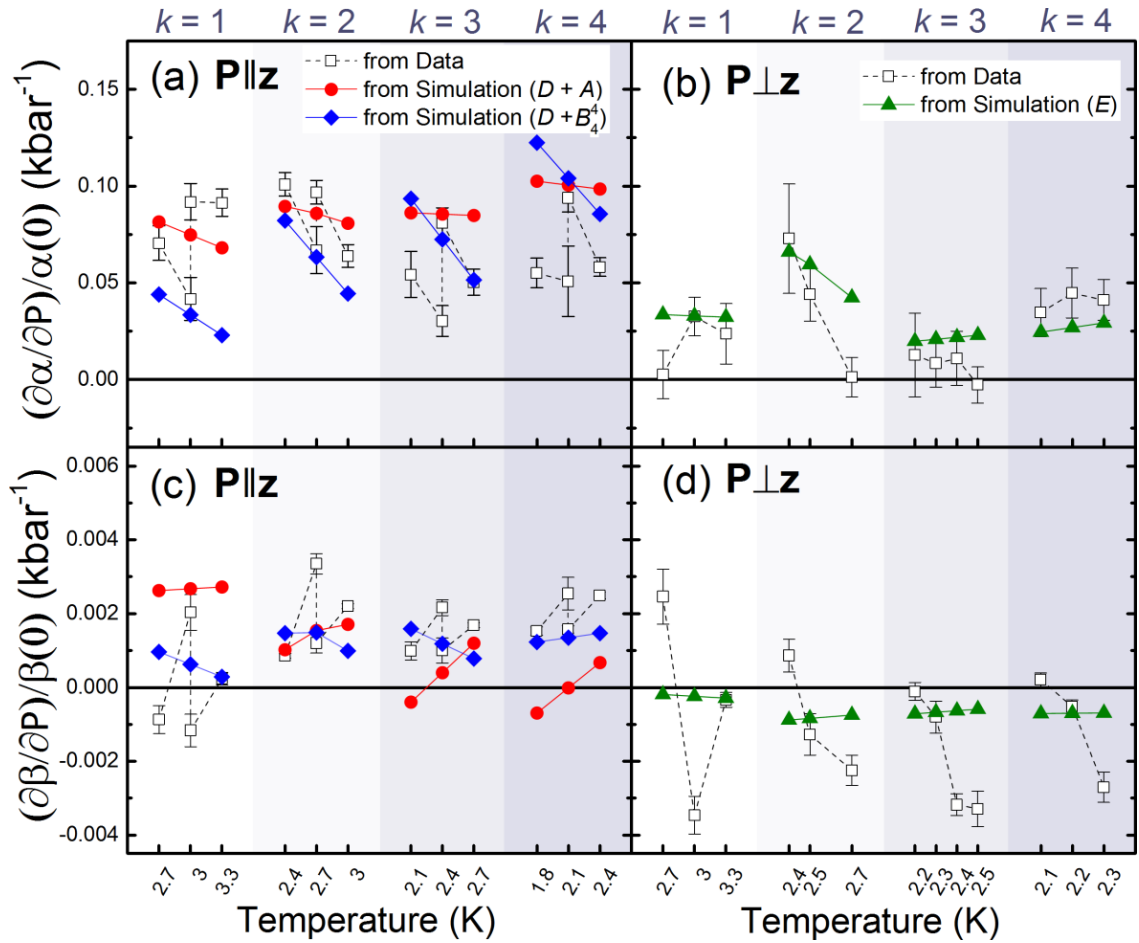

*Figure 5* Comparison of calculated and measured changes in the values of α and β. Colored circles/diamonds/triangles show the results of simulations (labelled in the legend by the anisotropy terms being modified in that simulation) while hollow squares are taken from experimental data, using the normalized slopes of the fits in the insets of *Figure 4* and fits to other similar data. The red circles correspond to a simulation of the relaxation rate using values generated by the two-state approximation. Some of the simulation data points represent extrapolations/interpolations of the observed (approximately linear) trends. The panels are sorted into different parameters and pressure configurations, with offsets to differentiate step number $k$ and temperature, as labelled above and below the plots. For the parallel configuration data, the duplicate entries at some temperature/step combinations represent data from two different sweep rates.

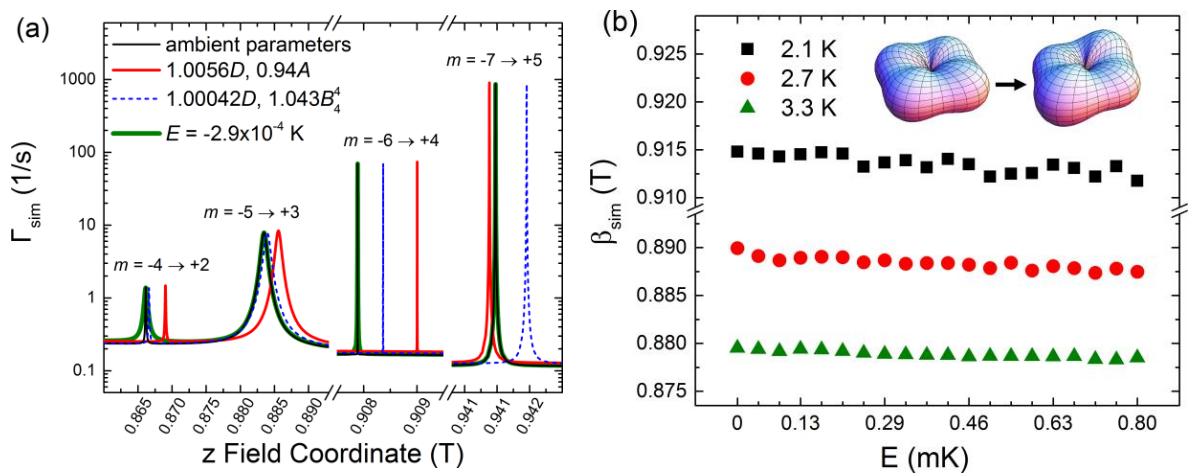

*Figure 6* (a) Numerical simulation of the relaxation rate for resonances involved in the $k = 2$ step, using the "ambient" anisotropy parameters as well as altered values, indicated in the legend, intended to mimic the effects of parallel and perpendicular pressures. The different transitions involved in each resonance are indicated above the set of peaks. (b) Estimated resonance position $\beta$ as a function of transverse anisotropy parameter $E$, extracted from the simulations; the inset illustrates an exaggerated distortion of a molecule's symmetry induced by the introduction of the rhombic $E$ term.

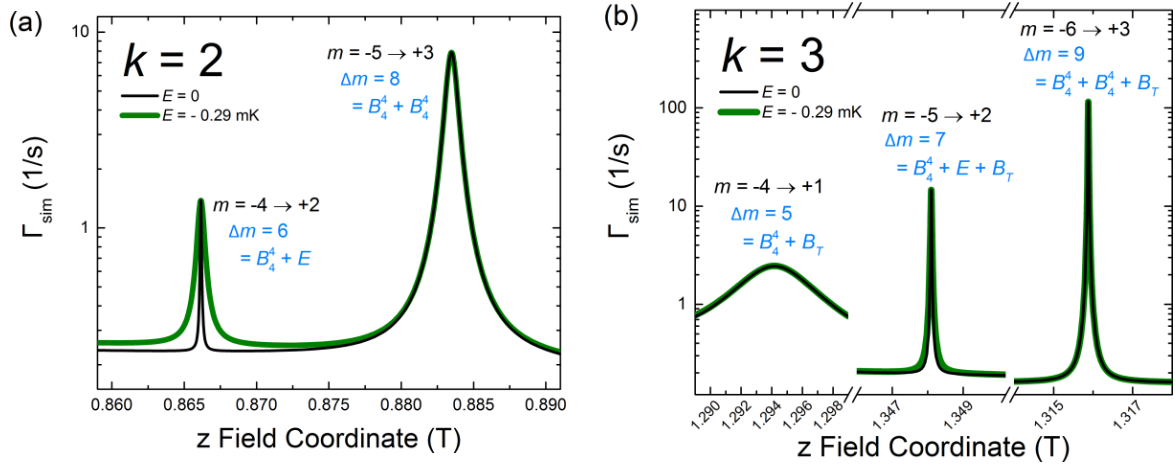

*Figure 7* Details of the calculated relaxation rate for "ambient" parameters (black) and for a system with an additional rhombic term with $E = -0.38$ mK (green) for several resonances at the $k = 2$ and 3 steps. The $k = 2$ data is also shown in *Figure 6*. The specific transitions are identified by the magnetic quantum numbers $m$ and $m'$ of the bare eigenstates at the level anticrossing, assuming a transition out of the metastable well. The change $\Delta m$ associated with each peak is identified, along with the anisotropy parameters involved (to leading order in perturbation theory) in mixing the states and enabling tunneling, highlighting the observed dependence on $E$.